# Random walk on Sierpiński-type multifractals

U. Bernert, K. Koepernik,
Fakultät für Physik und Geowissenschaften,
Universität Leipzig, Augustusplatz 10, D 04109 Leipzig,
bernert@tph100.physik.uni-leipzig.de

**Abstract**

A method is established which allows the calculation of the walk dimensions for Sierpiński-type multifractals. The multifractal scaling behaviour of the average time needed to cover a distance in the Sierpiński-type multifractals is shown. We calculate the Rényi dimensions for the average-time-multifractal and apply the $f(\alpha)$-formalism.

## 1 About the objects we are considering

The topic of our interest is the scaling behaviour of a certain set of geometrically constructed multifractals, the so-called Sierpiński-type multifractals. Strictly speaking, we generalize the set of Sierpiński-type fractals introducing a supplementary structure.

First some conventions. Any pure Sierpiński fractal is characterized by the embedding Euclidian dimension $d_e$ and the base $b$. Scaling the space coordinates with the factor $b$ yields a fractal point-set, which is indentical to the original fractal point-set. Further, we will call the simple geometric object representing the first step in the construction of a pure Sierpiński fractal an elementary generator of the base $b$.

There are two ways to construct Sierpiński fractals. On the one hand, we can manage it starting from a $d_e$-dimensional simplex. We divide all edge lines in $b$ parts. By connecting the division points with parallels to the edge lines we get a set of disjunct subcells, also simplices. Now we wipe out all simplices whose orientations differ from that of the original simplex. The number of remaining subcells in an elementary generator can be computed by the recursive formula:

$$n(b,d_e) = \begin{cases} 1 \quad \text{for} \quad d_e = 0 \\ \sum\limits_{k=1}^{b} n\left(k, d_e - 1\right) \quad \text{else}\,, \end{cases} \tag{1}$$

or their explicit equivalent:

$$n(b,d_e) = \binom{b + d_e - 1}{d_e}\,. \tag{2}$$

In the following we restrict ourselves to the cases $d_e \geq 2$. In the above picture we approach the Sierpiński fractal from the nearest higher-dimensional simplex. But before finishing the construction we are not able to define a reasonable model of walk. On the other hand, we can compose it as a gasket of one-dimensional line pieces. For every construction step we can define a walk model on the gasket. After an infinite number of repetitions both constructions yield the same fractal. But for our intentions the second picture will be better.

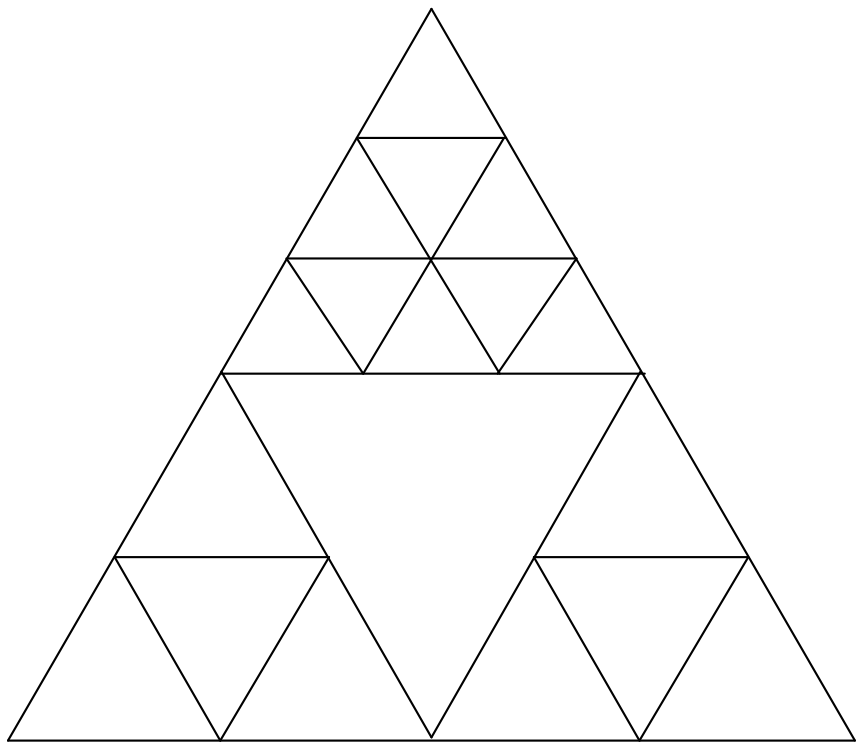

Figure 1: After the second construction step; here we have 2 generator types (bases 2 and 3) with the ratio of average numbers 2:1

Our Sierpiński-type fractals result by mixing some elementary generators of different bases (figure 1). After numbering all elementary generators we define a matrix $A$, whose elements $a_{ij}$ are the average number of $i$-type elementary generators in $j$-type elementary generators. Sometimes we will use the factorized form:

$$a_{ij} = g_{ij} n_j \qquad \text{with} \qquad n_j = n(b_j, d_e)\,. \tag{3}$$

Here $g_{ij}$ is the probability to find subcells with $i$-type elementary generators in the $j$-type parent elementary generator. Obviously we have

$$\sum_i g_{ij} = 1\,,\ \forall j\,. \tag{4}$$

The knowledge of the matrix $A$ does not completely determine our multifractal's geometrical structure. Also there is a stochastic element in it; we can not say anything about the exact distribution of subcells including different elementary generators in any parent elementary generator. The only thing we are sure is, that after a finite number $N$ of construction steps we can realize the distribution coefficients $g_{ij}$ as exactly as we wish. Thus, we will neglect in our calculations all contributions of the first $N$ construction steps and treat the multifractal as completely characterized by the given matrix $A$.

## 2 Rényi dimensions and $f(\alpha)$-formalism

To compute the Rényi dimensions $D_q$ and to apply the $f(\alpha)$-formalism we ought to mention some facts about the notion of dimensions for multifractal point-sets (see [1]). The so called local dimension $\alpha(\vec{r})$ describes the local scaling behaviour of some extensive (physical) property $p$ of our multifractal system. Marking all the points with the same value of $\alpha$ yields another fractal point-set. Its Hausdorff dimension is $f(\alpha)$. The Rényi dimensions are a set of dimensions parameterized by $q$. This set includes the Hausdorff dimension $D_0$, the information dimension $D_1$ and the correlation dimension $D_2$. These dimensions $D_q$ are so called global dimensions. That is, they characterize the whole multifractal point-set and do not have any local notion. But they are closely related to $\alpha$ and $f(\alpha)$ by the following formulas:

$$\alpha(q) = \frac{d}{dq}\Big((q-1)D_q\Big) \tag{5}$$

and:

$$D_q = \frac{1}{q-1}\Big(q\alpha(q) - f(\alpha(q))\Big)\,. \tag{6}$$

To get the Rényi dimensions of a given multifractal it was defined a partition function for any splitting $\{S_i\}$ of our multifractal into $K$ cells:

$$\Gamma(q,\tau,\{S_i\},l) = \sum_{i=1}^{K} \frac{p_i^q}{l_i^\tau}\,. \tag{7}$$

Here $p_i$ is the value of the considered property and $l_i$ is the edge-length of the $i$-th cell. In the limit of infinite small cells the sum above converges to $\Gamma(q,\tau)$ for any sequence of $\{S_i\}$. It was been shown that demanding $\Gamma(q,\tau)$ to be of order 1 (exactly speaking: $0 \leq \Gamma(q,\tau) \leq \infty$) leads to

$$\tau(q) = (q-1)D_q\,. \tag{8}$$

Thus we get

$$f(\alpha(q)) = q\alpha(q) - \tau(q)\,. \tag{9}$$

Now we will return to our multifractals. Here we use the set of our fractal's own subcells in the actual construction step as $\{S_i\}$. Just the question arises: What happens with the above sum when continuing the construction by one step? Obviously we must treat all contributions to the above sum from subcells including the same elementary generator (after the next construction step) in the same way. Therefore it seems useful splitting the sum $\Gamma(q,\tau)$ into the elements of a vector $\vec{\gamma}(q,\tau)$. Each element of this vector is directly related to one elementary generator type incoming in our multifractal's construction; it contains the contribution of all subcells including the same elementary generator type in the next construction step.

It is necessary to guarantee the right treatment of these partial sums in the further construction. It is the moment to use the matrix $A$. In each construction step it sorts the next sum in the correct way and places all terms in the right partial sum. The change of the summands due to the scaling will be managed by the two matrices $p_{ij}$ and $l_{ij}$. $p_{ij}$ is the amount of the extensive (physical) property of the subcell $i$ placed into the cell $j$ with

$$\sum_i g_{ij} n_j p_{ij} = 1\,,\ \forall j\,. \tag{10}$$

The $l_{ij} = l_j = 1/b_j$ are the edge-lengths of the $i$-type subcells in the parent cell of the type $j$ with the edge-length 1. Combining all this we get the transformation rule of $\vec{\gamma}(q,\tau)$ for the transition to the next construction step:

$$\vec{\gamma}^{(k+1)} = M\vec{\gamma}^{(k)} \qquad \text{with} \qquad M = (m_{ij}) = \left(a_{ij}\frac{p_{ij}^q}{l_{(i)j}^\tau}\right)\,. \tag{11}$$

The matrix $M$ contains all needed information to compute the multifractal dimensions. It only misses out the exact geometrical structure. That is, we know nothing about the concrete distribution of $\alpha(\vec{r})$. However, in $M$ we find all information about the global density of cells (in the limit: points) with the same value of $\alpha$ and that is sufficient for our calculations. Therefore it seems reasonable to name $M$ the "multifractal-generator(-matrix)". In our description "to be of order 1" is transferred from $\Gamma$ to the largest eigenvalue of $M$. This is equivalent to the demand for the existence of a nontrivial matrix $M^\infty$ with non-negative elements. To make sure that its largest eigenvalue must be equal to 1. It can be proven easily.

## 3 The walk dimension $d_w$

For the following we need the time scaling factors $\lambda$[1] for each type of elementary generators incoming in our multifractal's construction. Thus we start with proposing a method for their calculation. Its explicit derivation shall be omitted here. First we mark one of the corner points as the starting-point and the $d_e$ other corner points as end-points for any way of particles passing through our elementary generator's network. In our simple walking model the time needed to cover the distance between two nearest neighbor points of the elementary generator has the value 1 and there is no waiting time between two walking steps. Now we number all elementary generator points giving the same number to all that points in equivalent positions relating to the starting-point. Further, all end-points shall absorb the arriving particles completely. The next step is to write down the propagator containing all transition probabilities between nearest neighbor points. This suggests to use a vector-matrix-notation. The $i$-th element of the vector $\vec{X}^{(k)}$ used here contains the probability to find the particle located at the $i$-th point after the $k$-th step. The application of the propagator matrix $Q = (q_{ij})$ creates the probability distribution after the next step:

$$\vec{X}^{(k)} = Q^k \vec{X}^{(0)} \, . \tag{12}$$

The contribution to $Q$ from the end-points can be neglected due to their absorbing property. This makes $Q$ a little smaller. Further, it is reasonable to define the vector $\vec{X}$:

$$\vec{X} = \sum_{k=0}^{\infty} \vec{X}^{(k)} = \left( \sum_{k=0}^{\infty} Q^k \right) \vec{X}^{(0)} \, . \tag{13}$$

The absorbing end-points yield

$$Q^{\infty} = 0 \, . \tag{14}$$

Therefore we are allowed to write

$$\sum_{k=0}^{\infty} Q^k = \frac{1}{E - Q} = (E - Q)^{-1} \, . \tag{15}$$

Here $E$ is the identity matrix. We get

$$\vec{X} = \frac{1}{E - Q} \vec{X}^{(0)} \, . \tag{16}$$

If $h_i$ is the number of equivalent points specified by the same index $i$, the probability of passages through the link $ij$ in the $k$-th step is

$$H^{(k)}(i \leftrightarrow j) = \begin{cases} h_i x_i^{(k)} q_{ji} + h_j x_j^{(k)} q_{ij} & \text{for} \quad i \neq j \\ h_i x_i^{(k)} q_{ji} & \text{for} \quad i = j \, . \end{cases} \tag{17}$$

This yields the total (average!) number of passages through the link $ij$ counted over all possible ways:

$$H(i \leftrightarrow j) = \begin{cases} h_i x_i q_{ji} + h_j x_j q_{ij} & \text{for} \quad i \neq j \\ h_i x_i q_{ji} & \text{for} \quad i = j \, . \end{cases} \tag{18}$$

Now we get

$$\lambda = \frac{1}{2} \sum_{i,j} H(i \leftrightarrow j) = \sum_{i,j} q_{ij} x_j h_j \, . \tag{19}$$

[1]For any diffusion $\langle x^2 \rangle \propto t^{\frac{2}{d_w}}$ and $d_w = \frac{\ln \lambda}{\ln b}$

For having neglected the absorbing end-points as well in $\vec{X}^{(k)}$ and $\vec{X}$ as in the propagator matrix $Q$, we are allowed to simplify the above formula to

$$\lambda = \sum_i h_i x_i = \vec{H}^T \vec{X} = \vec{H}^T \frac{1}{E - Q} \vec{X}^{(0)} . \tag{20}$$

Back to our principal problem the question arises: Can we use the method described in the preceding section to compute the walk dimension $d_w$? Yes, we do as will be seen soon.

The following comments should enlighten our idea. In our model the walking particles leave behind massive traces with a constant one-dimensional density of mass. For the average walking path (in the complete multifractal), the geometrical structure built by all traces (weighted with the probability of their occurrence) exhibits the same scaling behaviour as the average walk time. This geometrical structure will be named "assigned multifractal". Its Hausdorff dimension is the walk dimension $d_w$ of our interest.

Just one can question the possibility to use the time scaling factors $\lambda$ found out by treating all elementary generators separately. And what happens with a particle returned to the starting-point without having reached any end-point, which starts then into another elementary generator? Concerning the calculation itself there are no difficulties; but the physical interpretation has to be discussed carefully.

As the same walk dimension $d_w$ is valid for all kinds of random walk models, we can postulate a certain one. This shall be a stationary process taking place in the network after $(k+1)$ construction steps. Further, it shall be an average process, what means – strictly spoken – it represents the average behaviour of particles. Then we consider the subcells of the $k$-th construction step as black boxes characterized by their substructures. After numbering all these black boxes, the only well-known thing is that per each time unit $N_j$ particles start from each corner point into the $j$-th black box. The symmetry of this idea causes the arrival of the same number $N_j$ of particles from the internal space of the $i$-th black box in each of their corner points. Introducing the argument of indistinguishability for particles located at the same point, we can suppose that the particles starting from any corner point into a related black box are the same who have just arrived from this black box. In this picture the particles can never leave their cells. This is exactly what we were looking for.

It remains the proof of renormalizability for the assigned multifractal. Therefore we refer to our method for calculating the time scaling factors $\lambda$. Including the absorbing end-points into $Q$, $\vec{X}^{(k)}$ and $\vec{X}$ and symmetrizing the problem by carrying out the sum over all possibilities to mark the starting-point we get the same total number of passages for all links in an elementary generator. Thus, the transition to the next construction step does not disturb our way of consideration.

Now let us remind the multifractal-generator-matrix $M$. It is known that we can choose any partition of the $p_{ij}$ satisfying the condition (10). As we would not complicate our calculations, we suppose the $p_{ij}$ does not depend on the index $i$. This yields the equipartition $p_j = 1/n_j$. Then the matrix $M$ becomes

$$m_{ij} = g_{ij} n_j \frac{\left(\frac{1}{n_j}\right)^q}{\left(\frac{1}{b_j}\right)^\tau} = g_{ij} \frac{b_j^\tau}{n_j^{q-1}} . \tag{21}$$

Here we must distinguish exactly the effect and the meaning of the several factors. The $g_{ij}$ sort the sum and guarantee the correct treatment of each partial sum and nothing more. The $b_j$ represent the length scaling behaviour and nothing else. And last but not least the $n_j$ stand for the "mass"-scaling behaviour and for nothing more.

For the assigned multifractal we can take over the $g_{ij}$ and the $b_j$ from the basic multifractal. However, the "mass"-scaling factors $n_j$ must be replaced by the "mass"-scaling factors of the assigned multifractal's elementary generators, the time scaling factors $\lambda_j$. Eventually we get

$$M_\lambda = (m_{ij})_\lambda = \left( g_{ij} \frac{b_j^\tau}{\lambda_j^{q-1}} \right)_\lambda . \tag{22}$$

In this case we have

$$p_j = \frac{1}{\lambda_j} . \tag{23}$$

The general form for any partition of the $p_{ij}$ is

$$M_\lambda = \left( g_{ij} \lambda_j \frac{p_{ij}^q}{l_j^\tau} \right)_\lambda = \left( g_{ij} \lambda_j p_{ij}^q b_j^\tau \right)_\lambda \qquad \text{with} \qquad \sum_i g_{ij} \lambda_j p_{ij} = 1 . \tag{24}$$

The cells of the basic multifractal have no direct meaning for the scaling properties of the assigned multifractal, the only direct relation is to the $g_{ij}$. Further, the assigned multifractal's Hausdorff dimension $d_w$ is larger than the embedding dimension $d_e$. This leads to the imagination of our assigned multifractal's proliferation into the $(d_e + 1)$-th dimension.

## 4 Example

We choose the two bases 2 and 3, the embedding dimension shall be 2. The numbers of subcells in the elementary generators are 3 and 6. Further, we set $g_{ij} = \frac{1}{2}$, $\forall i, j$ and $p_{ij} = p_j$. Thus we get

$$M = \begin{bmatrix} \frac{2^{\tau-1}}{3^{q-1}} & \frac{3^{\tau+1}}{6^q} \\ \frac{2^{\tau-1}}{3^{q-1}} & \frac{3^{\tau+1}}{6^q} \end{bmatrix} .$$

Demanding the largest eigenvalue to be equal 1 leads to the equation:

$$1 = \frac{2^{\tau-1}}{3^{q-1}} + \frac{3^{\tau+1}}{6^q} .$$

Solving the equation yields the Hausdorff dimension $D_0 = 1.613\ldots$ and the figure 2. Applying the $f(\alpha)$-formalism we get the figure 3. Just we come to the time scaling factors. Therefore we need

$$Q_1 = \begin{bmatrix} 0 & \frac{1}{2} & 0 \\ \frac{1}{2} & \frac{1}{4} & \frac{1}{4} \\ 0 & \frac{1}{2} & 0 \end{bmatrix} \qquad \text{and} \qquad Q_2 = \begin{bmatrix} 0 & \frac{1}{2} & 0 & 0 & 0 \\ \frac{1}{2} & \frac{1}{4} & \frac{1}{6} & \frac{1}{4} & 0 \\ 0 & \frac{1}{2} & 0 & \frac{1}{2} & \frac{1}{2} \\ 0 & \frac{1}{4} & \frac{1}{6} & 0 & \frac{1}{4} \\ 0 & 0 & \frac{1}{6} & \frac{1}{4} & \frac{1}{4} \end{bmatrix} .$$

Further, we are given

$$\vec{H}_1 = \begin{bmatrix} 1 \\ 2 \\ 1 \end{bmatrix} \qquad \text{and} \qquad \vec{H}_2 = \begin{bmatrix} 1 \\ 2 \\ 1 \\ 2 \\ 2 \end{bmatrix} .$$

With $\vec{x}_j^{(0)} = \delta_{j,1}$ for all bases we get

$$\lambda_1 = 5 \qquad \text{and} \qquad \lambda_2 = \frac{90}{7} .$$

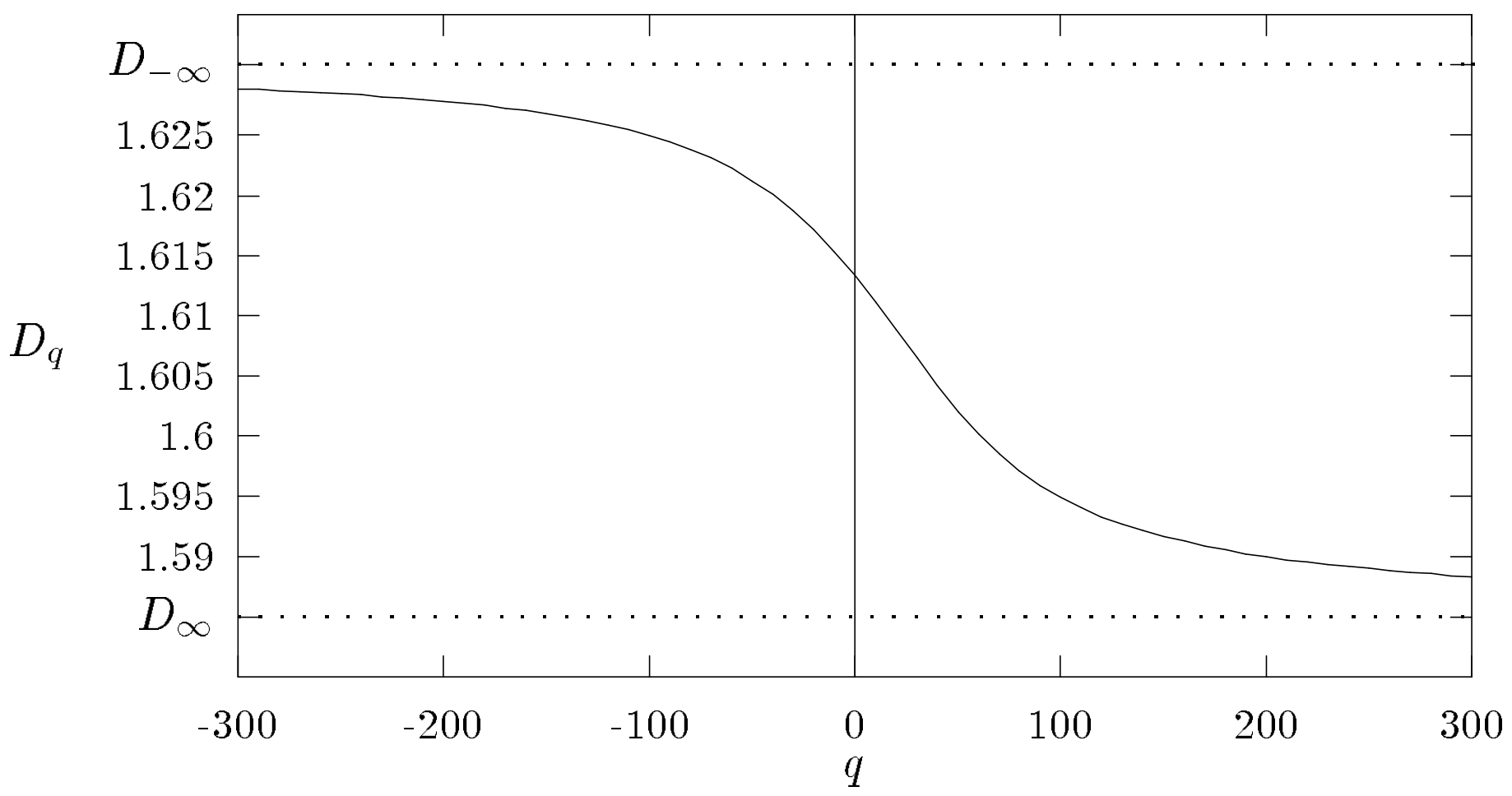

Figure 2: The Rényi dimension $D_q$

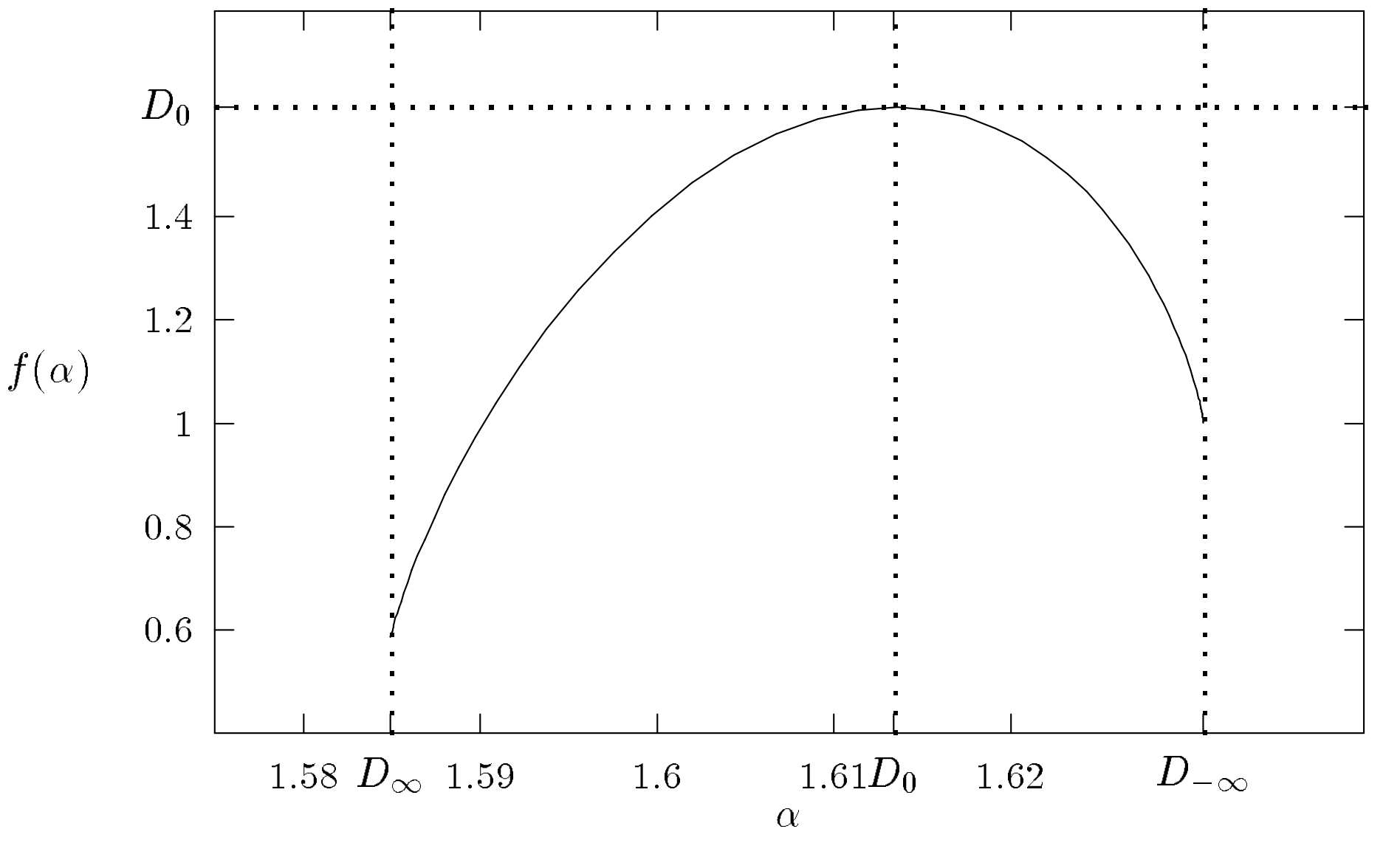

Figure 3: The global dimension $f(\alpha)$ vs. the local dimension $\alpha$

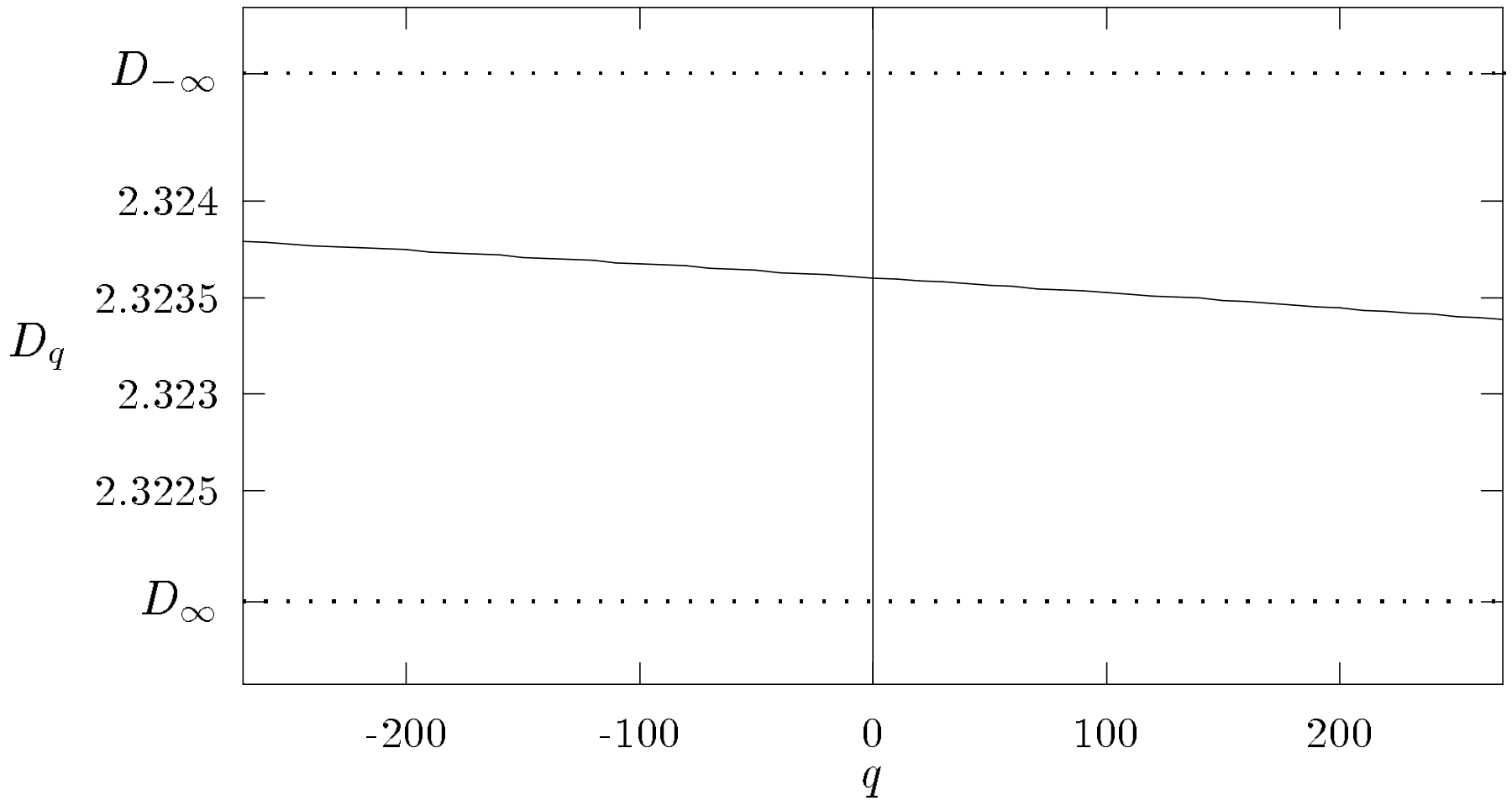

Figure 4: The Rényi dimension $D_q$ for the assigned multifractal

Now we write down the matrix $M_\lambda$ for the assigned multifractal:

$$M_\lambda = \frac{1}{2} \begin{bmatrix} \frac{2^\tau}{5^{q-1}} & \frac{3^\tau}{\left(\frac{90}{7}\right)^{q-1}} \\ \frac{2^\tau}{5^{q-1}} & \frac{3^\tau}{\left(\frac{90}{7}\right)^{q-1}} \end{bmatrix} .$$

Demanding the largest eigenvalue to be equal 1 leads to the equation:

$$1 = \frac{1}{2} \left( \frac{2^\tau}{5^{q-1}} + \frac{3^\tau}{\left(\frac{90}{7}\right)^{q-1}} \right) .$$

Solving the equation yields the walk dimension $d_w = 2.323\ldots$ and the figure 4. There we have $D_{-\infty} = d_w(b = 3)$ and $D_\infty = d_w(b = 2)$. Applying the $f(\alpha)$-formalism we get the figure 5.

# 5 Conclusion

Obviously the calculation of the walk dimension essential for every kind of diffusion works with the simplest walk model. The model of the average-walk-time-multifractal has proven to be good. For the dimensions there is the relation: $D_0 < d_e < d_w$. Our method can be seen as a generalized version of the renomalization method already used to calculate walk dimensions.

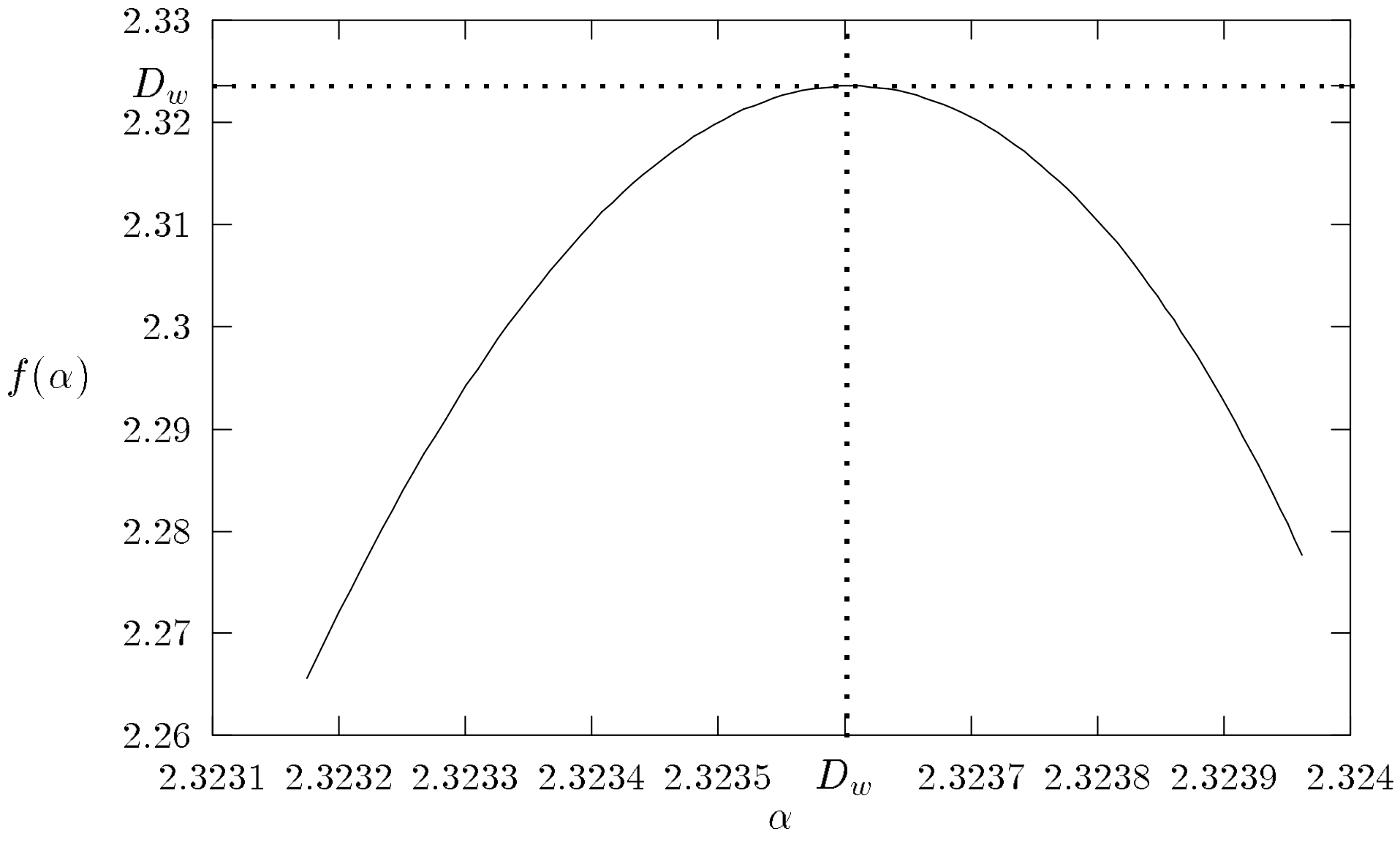

Figure 5: The global dimension $f(\alpha)$ vs. the local dimension $\alpha$ for the assigned multifractal